\documentclass[preprint,showpacs,groupedaddress,amsmath,amssymb]{revtex4}
\usepackage{graphicx}
\usepackage{dcolumn}
\usepackage{multirow}
\usepackage{amsfonts}

\newcommand{\beq}{\begin{equation}}
\newcommand{\eeq}{\end{equation}}
\newcommand{\bea}{\begin{align}}
\newcommand{\eea}{\end{align}}
\newcommand{\epsi}{\varepsilon}
\newcommand{\Cth}{\cos \theta}
\newcommand{\Sth}{\sin \theta}
\newcommand{\del}{\nabla}

\newcommand{\PI}{\overleftrightarrow{\mathbf{\Pi}}}
\newcommand{\WT}{\overleftrightarrow{\mathbf{W}}}
\newcommand{\vphi}{\varphi}
\newcommand{\vtheta}{\vartheta}

\newcommand{\half}{\frac{1}{2}}
\newcommand{\wt}[1]{\widetilde{#1}}

\makeatletter
\def\Tbar{\mathchoice
   {\TTbar\displaystyle\textstyle{-}}%
   {\TTbar\textstyle\scriptstyle{-}}%
   {\TTbar\scriptstyle\scriptscriptstyle{-}}%
   {\TTbar\scriptscriptstyle\scriptscriptstyle{-}}%
   \!T}
\def\TTbar#1#2#3{{\setbox0=\hbox{$#1{#2#3}{\mathrm{T}}$}
     \raise2\p@\vbox{\hbox{$#2#3$}}\kern-.35\wd0}}
\makeatother

\begin{document}

\title{Viscosity and confinement in magnetized plasma}


\author{Robert W. Johnson}
\email[]{rob.johnson@gatech.edu}
\affiliation{Fusion Research Center, Georgia Institute of Technology, Neely Building, Atlanta, GA 30332, USA}


\date{March 15, 2007.  Revised: October 10, 2007.}

\begin{abstract}
An alternative to the Braginskii decomposition is proposed, one rooted in treating the viscosity as a scalar quantity in a coordinate-free representation.  With appropriate application to the rate-of-shear tensor, one may solve the neoclassical force density equations for its undetermined velocity dependence, as well as the radial and poloidal profiles mentioned in~[R. W. Johnson, Phys. Plasmas, {\it under review}], using an improved poloidal expansion.  The pseudoplastic behavior of magnetized plasma is again obtained, and the high viscosity solution is determined to be physical.  A clear relationship between confinement mode and viscosity is observed, indicating a physical origin for transport barriers, pedestals, and other phenomena.  The gyroviscous contribution is found to be an effect on the order of one one-thousandth of one percent of the dominant collisional viscosity.
\end{abstract}


\maketitle


\section{Introduction}
Recognizing that quantities which may be written in a coordinate-independent manner retain a physical unity that transcends ones choice of representation, but also realizing that the magnetization of a plasma introduces potentially complicated effects, we make further improvements upon the standard theory of plasma viscosity~\cite{brag-1965,shaingetal-1985,stacandsig-1985} which was initiated in Reference~\cite{rwj-2007pa}.  We isolate the collisional and gyroviscous contributions to the scalar viscosity, leaving only its undetermined velocity dependence, the main ion velocities, and the density coefficients as the free parameters for which to solve the equations.  We also realize the proper poloidal expansion for physical quantities in our chosen geometry.  We compare the velocity dependence of the toroidal and poloidal viscosities for a variety of confinement regimes: L-mode, L-mode with Internal Transport Barrier, H-mode, and Quiescent H-mode, and observe that the quality of confinement is predicted by the slope of the viscosity-velocity dependence.  We observe that the internal transport barrier discharge has a toroidal pseudoplasticity much like the other L-mode discharge and a poloidal pseudoplasticity more like the H-mode shots, suggesting that viscosity is a determining factor of confinement mode.  We compare the gyroviscous and the collisional viscosity profiles for our collection of discharges and find that the collisional viscosity is the dominant effect by a factor on the order of $\mathcal{O}(10^5)$ over the majority of the profile, with slightly more gyroviscous contribution as the edge is approached and the various poloidal coefficients approach unity in magnitude.  We conclude by summarizing and by considering the implications of pseudoplastic plasma on fusion tokamak research, as well as on other aspects of plasma physics.

\section{Further Development}
\subsection{Poloidal Expansion}
We start from the same continuity and force density equations as in Reference~\cite{rwj-2007pa}, except for the attention which we will pay to the shear viscosity, and we use the same notation.  We account for the geometrical effects on physical quantities by first applying the flux surface measure factor to the denominator of an expanded quantity, much like the manner in which the magnetic field $\mathbf{B}=\mathbf{B}^0/(1+\epsi\Cth)$ is expressed, then taking the first order Fourier expansion, {\it eg}:
\beq \label{eqn:expan}
n_j (\theta) = n_j^0 \left( 1 + \epsi\wt{n}_j^c \Cth + \epsi\wt{n}_j^s \Sth \right) / (1+\epsi\Cth) \; ,
\eeq
where $\epsi\wt{n}_j^{c/s} = n_j^{c/s}$ has unit normalization, still neglecting the radial derivatives of the poloidal coefficients.  With this expansion, the unity, $\cos$, and $\sin$ flux surface moments of an expanded quantity pick up only their intended term:
\begin{align} \label{eqn:fsavn}
\left< n_j \right> & \equiv \frac{1}{2\pi} \oint_0^{2\pi} d\theta \left( 1+\epsi\Cth \right) n_j (\theta) = n_j^0 \; , & \\
\left< n_j \right>_C & \equiv  \frac{1}{2\pi} \oint_0^{2\pi} d\theta \Cth \left( 1+\epsi\Cth \right) n_j (\theta) = \half \epsi n_j^0 \wt{n}_j^c \; , \mathrm{and} & \\
\left< n_j \right>_S & \equiv  \frac{1}{2\pi} \oint_0^{2\pi} d\theta \Sth \left( 1+\epsi\Cth \right) n_j (\theta) = \half \epsi n_j^0 \wt{n}_j^s \; . & 
\end{align}
We note that intermediate results indicate that the choice of expansion has little effect on the solution near the core of the plasma but does play a significant role as $\epsi \equiv r/R_0 $ increases, rendering moot calculations near the edge based on the previous expansion.

\subsection{Viscosity}
Returning to the definition of the viscous stress tensor (for solenoidal flows), as given in a coordinate-free notation:
\beq \label{eqn:visc}
\PI \equiv \eta \; \WT \; ,
\eeq
where $\WT$ is the full shear tensor, we note that Braginskii~\cite{brag-1965} decomposes both the scalar viscosity and the shear tensor and then applies certain portions of the decomposed viscosity to certain portions of the decomposed shear, to produce the traditional neoclassical viscous stress tensor.  Without experimental verification, however, this procedure remains to be validated.  Indeed, unpublished results from the effort behind References~\cite{rwj-2007pa,frc-pop-2006} indicate that the neoclassical model cannot account for the observed velocity profiles in a predictive fashion.

We take the scalar viscosity ($\forall$ species $j$) to be the sum of the gyroviscous and the collisional contributions, {\it ie}
\beq
\eta_j = \eta_j^{gyro} + \eta_j^{coll} \; ,
\eeq
where the gyroviscosity and the collisional viscosity are those of Braginskii~\cite{brag-1965}.  (Using Shaing's plateau enhanced collision frequency~\cite{shaingetal-1985} did not produce believable results in this analysis, but more investigation is needed before it can be ruled out entirely.)  The shear tensor for species $j$ is taken as~\cite{stacandsig-1985}:
\beq
W_{\alpha \beta}^j \equiv \partial_\alpha V_\beta^j + \partial_\beta V_\alpha^j - \frac{2}{3} \delta_{\alpha \beta} \del \cdot \mathbf{V}^j \; ,
\eeq
where $\delta_{\alpha \beta}$ is the Kronecker delta, and is evaluated directly.  (Note that the species index can be either a sub- or super-script for clarity of notation.)

Here we depart from Braginskii's model---while Equation~(\ref{eqn:visc}) is correct for a normal fluid, as well as for an unmagnetized plasma, the presence of a magnetic field has no small impact on the behavior of the plasma.  The gyromotion of the ions produces two effects:  firstly, all the ions on a particular flux-surface execute their gyromotion in lockstep, thus the collisional viscosity should be applied only to those terms in the shear tensor without a derivative across flux-surfaces; secondly, the ions on adjacent flux-surfaces collide at the gyrofrequency, thus the gyroviscosity should be applied only to those terms containing the derivative across flux-surfaces.  We write our model's stress tensor as
\beq \label{eqn:newpi}
\PI_j \equiv \PI_j^{gyro} + \PI_j^{coll} = \eta_j^{gyro} \WT_j^{gyro} + \eta_j^{coll} \WT_j^{coll} \; .
\eeq
In order to account for the as-yet undetermined velocity dependence, we multiply the viscosities $\eta_j$ in the $\alpha$ equation by a free parameter, $\Gamma_\alpha^j$, for which we will solve the unity, $\cos$, and $\sin$ flux surface moment equations.  To keep our count of equations and unknowns equal, we construct both $\Gamma_\phi^j$ from a single free parameter via $\Gamma_\phi^j = \Gamma_\phi^0 \left< P_j^0 \right>_j / P_j^0$.

\section{Application and Analysis}
\subsection{Solution Procedure and Input Profiles}
The astute reader will notice that the toroidal unity moment equation now depends explicitly on the poloidal coefficients and will no longer tolerate our previous trick of isolating it from the others---we will have to solve the full system of equations at each $\rho$.  To solve our system, we input profiles for density, temperature, poloidal magnetic field (intrinsically related to the toroidal current, hence the toroidal electric field as well), toroidal momentum injection, and the rotation profiles for at least one species of ion.  To analyze shots from DIII-D~\cite{diiid-2002}, we retrieve the necessary magnetic geometry and density and temperature profiles from EFIT~\cite{Lao:1985mw}, the rotation profiles for Carbon-6 come from GAProfiles~\cite{gaprof-2000}, and the momentum injection is calculated by NBEAMS~\cite{nbeams-1992}.  The toroidal electric field is evaluated from the plasma current profile using Braginskii's~\cite{brag-1965} resistivity and Wesson's~\cite{tokamaks-2004} Coulomb logarithm.  We neglect the electron and radial flows in the following analysis.
Our selection of discharges now encompasses a wider variety of confinement regimes, one shot from each of what we will nominate as L, L-ITB, H, and QH modes; see Table~\ref{tab:shotdata} for the shot numbers and times we analyze.  All the shots use corrected toroidal velocities~\cite{solomonetal-pop-2006}, and the QH shot also has corrected poloidal velocities~\cite{solomon-words}.  The input profiles and radial electric field profiles are given in the Appendix.

\subsection{L Mode Shot 98777}
Let us first analyze shot 98777 in detail.  As before, we find two classes of solution, one with low viscosity and one with high viscosity, displayed in Figures~1 and 2,~
but this time we will have a criterion for discrimination.  As we have applied independent free parameters to the toroidal and poloidal components of $\del \cdot \PI$, the solution is allowed to come to rest at independent values for the viscosity associated with flows in the toroidal and poloidal directions, but as our theory is rooted in the scalar nature of viscosity, one should demand that the physical solution have essentially equal viscosities associated with both of these flows.  Comparing subfigures (c) and (d) of Figures~1 and 2,~
 we see that the high viscosity solution has much better agreement between the toroidal and poloidal viscosities, with the velocity dependence accounting for the difference.  The viscosities for the deuterium are essentially the same line.  The low viscosity solution displays a marked difference in the viscosities associated with the orthogonal flows, and is discounted as unphysical despite some readers' preference for nearly equivalent toroidal velocities between the species.  We also note the appearance of extreme pseudoplastic or glassy behavior in carbon's high viscosity solution, confirming our earlier suspicions~\cite{rwj-2007pa}.  That the previous model's high viscosity solution is so different than the current one we can only attribute to the previous model's improper treatment of viscosity in general (we feel that perhaps a relative - sign should have been applied to the toroidal viscosities in~\cite{rwj-2007pa}, which is now provided by the species-dependent poloidal coefficients).

We can investigate the poloidal coefficient profiles for these two solutions, Figures~3 and 4.~
Half of the poloidal coefficients have roughly equal magnitude values for the two solutions: the density coefficients and the poloidal velocity $\sin$ coefficients, with opposite signs for the $\sin$ coefficients (which are easier to observe in the online figures).  The variation in deuterium's toroidal velocity $\sin$ coefficient is accounted for by the variation in the velocity profiles.  The intriguing difference is in subfigures (d) and (f)---the low viscosity solution has nearly equivalent toroidal velocity $\cos$ coefficients and poloidal velocity $\cos$ coefficients with the same shape and a nearly constant factor between the species, while the high viscosity solution has deuterium's poloidal velocity $\cos$ coefficient of the same shape as its toroidal velocity $\cos$ coefficients and has carbon's poloidal velocity $\cos$ coefficient as in the low viscosity solution.  What these variations signify has yet to be determined.

We note in passing that the vanishing of the poloidal electric field (when radial electron flows are neglected) established in Reference~\cite{rwj-2007pa} has consequences beyond the vanishing of the poloidal coefficients of the radial electric field---that the electron density $\sin$ coefficient vanishes is a direct consequence, and empirically we find that the electron density $\cos$ coefficient vanishes as well.  We display these vanishing quantities for the high viscosity solution for both the new and old models (which stored only one species' determination of $E_r^{c/s}$) in Figure~5.

\subsection{Other Confinement Modes}
The velocity and viscosity profiles for the high viscosity solution for the remainder of the confinement modes are displayed in Figures~6 through 8.~
The poloidal coefficient solution profiles are given in the Appendix.  We note the fairly consistent agreement between the toroidal and poloidal viscosities and attribute the differences between the various viscosities as a velocity dependent effect driving the values away from a single, common scalar value.  Pseudoplasticity is observed across the board, with an interesting suppression of glassy behavior on the magnetic axis, $\rho = 0$.  The glassy phenomena is found to occur whenever a velocity approaches zero away from the magnetic axis.  Glassy peaks are observed in the poloidal viscosity for all the shots, and in the toroidal viscosity for shot 99411.  These glassy peaks near the edge might explain the formation of the edge pedestal and the edge transport barrier, and the relatively sudden change in toroidal viscosity in the middle of the profile of the L-ITB shot might explain the internal transport barrier.

\subsection{Interpretation}
We take the (log of the absolute value of the) toroidal and poloidal viscosities divided by the energy density (in other words, $\Gamma \eta / n \Tbar = \Gamma \tau$, which is neither the dynamic nor the kinematic viscosity, yet appropriate for the current analysis) and plot them against the velocities for our collection of confinement modes.  This time we combine the deuterium and carbon values and separate the toroidal and poloidal values, as displayed in Figure~9.~
We observe both pseudoplastic behavior and the suppression of pseudoplasticity (as evidenced by a horizontal relation), particularly in the poloidal viscosity values coming from the inner half-radius of $\rho$.  We take a linear fit of the combined deuterium and carbon data, over the whole profile for the toroidal relations and over the outer half-radius for the poloidal relations, and label the fitted lines as in the figure.  The values for the pseudoplastic slopes $S_\alpha$ are given in Table~\ref{tab:tautab}, but as the axes are all on equal scales (relatively, for the velocity axes), one may trust ones eye to compare the slopes directly on the figure.  We observe that the L and L-ITB shots display a greater pseudoplasticity than the H and QH shots, and that the values for the toroidal pseudoplasticity seem roughly consistent for both shots of L and H mode confinement.  The L mode is found to have a greater poloidal pseudoplasticity than the others, and we note that the L-ITB has a poloidal pseudoplsaticity more like the H and QH modes'.  It seems that a measure of the toroidal and poloidal pseudoplasticities could be used to define a quality of confinement
\beq
Q \equiv - \left( S_\vphi + S_\vtheta \right) \; ,
\eeq
that is on a quantitative scale rather than being a qualitative description, and our values for that quantity are in the last column of Table~\ref{tab:tautab}.  The lower the value of $Q$, the better the quality of confinement.

We may also isolate the relative contributions of the gyroviscous and the collisional viscosities, Figure~10.~
We observe that there is little variation between the species' toroidal viscosities and much more variation in the poloidal viscosities, with carbon having the greater viscosity.  We also observe (note the logarithmic scale) that gyroviscosity is dominated by the collisional viscosity by a factor of order $\mathcal{O}(10^5)$, hence accounts for one one-thousandth of one percent of the total viscosity present.

\section{Conclusion}
By proposing an alternative to Braginskii's decomposition of the stress tensor, we determine the pseudoplasticity of magnetized plasma in a fusion tokamak.  We observe clear relations between pseudoplasticity and confinement mode and suggest a quantitative measure of the quality of confinement based thereon.  The phenomena of glassy plasma persists and is suggested as the physical explanation for a variety of phenomena in magnetized plasma, including but not limited to pedestal effects and transport barriers.  The implications for other systems of magnetized plasma, both natural and man-made, and for other types of plasma, such as the quark-gluon plasma and the gravitational plasma models of galaxies, clearly need to be investigated by workers in a variety of fields to determine the importance of pseudoplastic viscosity on our understanding of plasma phenomena.  We end with a conjecture on cosmology---when the early universe went through a phase of being electronic plasma, then pseudoplastic viscosity would imply that as flow velocities decreased the viscosity would increase until neutralization occurred, impacting the standard comsological model~\cite{Guth-1981,Liddle-1999mq,Ghosh-1999xm} in unexpected ways, as well as perhaps seeding the initial density fluctuations from which galaxies were born.

\begin{acknowledgments}
The author gratefully acknowledges the helpful comments of Robert W. Johnson, Sr., and the contributions of the DIII-D Team who made the measurements presented in this paper, making use of data provided by the Atomic Data and Atomic Structure (ADAS) database---the originating developer of ADAS is the JET Joint Undertaking.  This work is privately funded.
\end{acknowledgments}

\appendix*
\section{Additional Figures}

\subsection{Input Profiles}

\subsection{Poloidal Coefficients}

\newpage


\begin{thebibliography}{15}
\expandafter\ifx\csname natexlab\endcsname\relax\def\natexlab#1{#1}\fi
\expandafter\ifx\csname bibnamefont\endcsname\relax
  \def\bibnamefont#1{#1}\fi
\expandafter\ifx\csname bibfnamefont\endcsname\relax
  \def\bibfnamefont#1{#1}\fi
\expandafter\ifx\csname citenamefont\endcsname\relax
  \def\citenamefont#1{#1}\fi
\expandafter\ifx\csname url\endcsname\relax
  \def\url#1{\texttt{#1}}\fi
\expandafter\ifx\csname urlprefix\endcsname\relax\def\urlprefix{URL }\fi
\providecommand{\bibinfo}[2]{#2}
\providecommand{\eprint}[2][]{\url{#2}}

\bibitem[{\citenamefont{{Braginskii}}(1965)}]{brag-1965}
\bibinfo{author}{\bibfnamefont{S.~I.} \bibnamefont{{Braginskii}}}, in
  \emph{\bibinfo{booktitle}{Review of Plasma Physics}}, edited by
  \bibinfo{editor}{\bibfnamefont{M.}~\bibnamefont{Leontovich}}
  (\bibinfo{publisher}{Consultants Bureau}, \bibinfo{address}{New York,
  U.S.A.}, \bibinfo{year}{1965}), vol.~\bibinfo{volume}{1} of
  \emph{\bibinfo{series}{Review of Plasma Physics}}, pp.
  \bibinfo{pages}{201--311}.

\bibitem[{\citenamefont{{Stacey} et~al.}(1985)\citenamefont{{Stacey}, {Bailey},
  {Sigmar}, and {Shaing}}}]{shaingetal-1985}
\bibinfo{author}{\bibfnamefont{W.~M.} \bibnamefont{{Stacey}}},
  \bibinfo{author}{\bibfnamefont{A.~W.} \bibnamefont{{Bailey}}},
  \bibinfo{author}{\bibfnamefont{D.~J.} \bibnamefont{{Sigmar}}},
  \bibnamefont{and} \bibinfo{author}{\bibfnamefont{K.~C.}
  \bibnamefont{{Shaing}}}, \bibinfo{journal}{Nucl. Fusion}
  \textbf{\bibinfo{volume}{25}}, \bibinfo{pages}{463} (\bibinfo{year}{1985}).

\bibitem[{\citenamefont{{Stacey} and {Sigmar}}(1985)}]{stacandsig-1985}
\bibinfo{author}{\bibfnamefont{W.~M.} \bibnamefont{{Stacey}}} \bibnamefont{and}
  \bibinfo{author}{\bibfnamefont{D.~J.} \bibnamefont{{Sigmar}}},
  \bibinfo{journal}{Phys. Fluids} \textbf{\bibinfo{volume}{28}},
  \bibinfo{pages}{2800} (\bibinfo{year}{1985}).

\bibitem[{\citenamefont{{Johnson}}(2007)}]{rwj-2007pa}
\bibinfo{author}{\bibfnamefont{R.~W.} \bibnamefont{{Johnson}}},
  \bibinfo{journal}{Phys. Plasmas}  (\bibinfo{year}{2007}),
  \bibinfo{note}{under review}, \eprint{arXiv:0710.2657v1 [physics.plasm-ph]}.

\bibitem[{\citenamefont{{Stacey} et~al.}(2006)\citenamefont{{Stacey},
  {Johnson}, and {Mandrekas}}}]{frc-pop-2006}
\bibinfo{author}{\bibfnamefont{W.~M.} \bibnamefont{{Stacey}}},
  \bibinfo{author}{\bibfnamefont{R.~W.} \bibnamefont{{Johnson}}},
  \bibnamefont{and}
  \bibinfo{author}{\bibfnamefont{J.}~\bibnamefont{{Mandrekas}}},
  \bibinfo{journal}{Phys. Plasmas} \textbf{\bibinfo{volume}{13}}
  (\bibinfo{year}{2006}).

\bibitem[{\citenamefont{{Luxon}}(2002)}]{diiid-2002}
\bibinfo{author}{\bibfnamefont{J.~L.} \bibnamefont{{Luxon}}},
  \bibinfo{journal}{Nucl. Fusion} \textbf{\bibinfo{volume}{42}}
  (\bibinfo{year}{2002}).

\bibitem[{\citenamefont{Lao et~al.}(1985)\citenamefont{Lao, St.~John,
  Stambaugh, Kellman, and Pfeiffer}}]{Lao:1985mw}
\bibinfo{author}{\bibfnamefont{L.}~\bibnamefont{Lao}},
  \bibinfo{author}{\bibfnamefont{H.}~\bibnamefont{St.~John}},
  \bibinfo{author}{\bibfnamefont{R.}~\bibnamefont{Stambaugh}},
  \bibinfo{author}{\bibfnamefont{A.}~\bibnamefont{Kellman}}, \bibnamefont{and}
  \bibinfo{author}{\bibfnamefont{W.}~\bibnamefont{Pfeiffer}},
  \bibinfo{journal}{Nucl. Fusion} \textbf{\bibinfo{volume}{25}},
  \bibinfo{pages}{1611} (\bibinfo{year}{1985}).

\bibitem[{\citenamefont{{Zeng} et~al.}(2000)\citenamefont{{Zeng}, {Doyle},
  {Peebles}, and {Luce}}}]{gaprof-2000}
\bibinfo{author}{\bibfnamefont{L.}~\bibnamefont{{Zeng}}},
  \bibinfo{author}{\bibfnamefont{E.~J.} \bibnamefont{{Doyle}}},
  \bibinfo{author}{\bibfnamefont{W.~A.} \bibnamefont{{Peebles}}},
  \bibnamefont{and} \bibinfo{author}{\bibfnamefont{T.~C.}
  \bibnamefont{{Luce}}}, \bibinfo{journal}{APS Meeting Abstracts}, \bibinfo{misc}{American Physical Society, 42nd Annual Meeting of the APS Division of Plasma Physics combined with the 10th International Congress on Plasma Physics October 23 - 27, 2000 Québec City, Canada Meeting ID: DPP00, abstract \#NP1.102}
(\bibinfo{year}{2000}).

\bibitem[{\citenamefont{{Mandrekas}}(1992)}]{nbeams-1992}
\bibinfo{author}{\bibfnamefont{J.}~\bibnamefont{{Mandrekas}}},
  \bibinfo{type}{Tech. Rep.}, \bibinfo{institution}{Georgia Institute of
  Technology} (\bibinfo{year}{1992}).

\bibitem[{\citenamefont{{Wesson}}(2004)}]{tokamaks-2004}
\bibinfo{author}{\bibfnamefont{J.}~\bibnamefont{{Wesson}}},
  \emph{\bibinfo{title}{Tokamaks}} (\bibinfo{publisher}{Cambridge University
  Press}, \bibinfo{address}{Cambridge, England}, \bibinfo{year}{2004}),
  \bibinfo{edition}{3rd} ed.

\bibitem[{\citenamefont{{Solomon} et~al.}(2006)\citenamefont{{Solomon},
  {Burrell}, {Andre}, {Baylor}, {Budny}, {Gohil}, {Groebner}, {Holcomb},
  {Houlberg}, and {Wade}}}]{solomonetal-pop-2006}
\bibinfo{author}{\bibfnamefont{W.~M.} \bibnamefont{{Solomon}}},
  \bibinfo{author}{\bibfnamefont{K.~H.} \bibnamefont{{Burrell}}},
  \bibinfo{author}{\bibfnamefont{R.}~\bibnamefont{{Andre}}},
  \bibinfo{author}{\bibfnamefont{L.~R.} \bibnamefont{{Baylor}}},
  \bibinfo{author}{\bibfnamefont{R.}~\bibnamefont{{Budny}}},
  \bibinfo{author}{\bibfnamefont{P.}~\bibnamefont{{Gohil}}},
  \bibinfo{author}{\bibfnamefont{R.~J.} \bibnamefont{{Groebner}}},
  \bibinfo{author}{\bibfnamefont{C.~T.} \bibnamefont{{Holcomb}}},
  \bibinfo{author}{\bibfnamefont{W.~A.} \bibnamefont{{Houlberg}}},
  \bibnamefont{and} \bibinfo{author}{\bibfnamefont{M.~R.}
  \bibnamefont{{Wade}}}, \bibinfo{journal}{Phys. Plasmas}
  \textbf{\bibinfo{volume}{13}} (\bibinfo{year}{2006}).

\bibitem[{\citenamefont{{Solomon}}(2006)}]{solomon-words}
\bibinfo{author}{\bibfnamefont{W.~M.} \bibnamefont{{Solomon}}},
  \emph{\bibinfo{title}{private communication}} (\bibinfo{year}{2006}).

\bibitem[{\citenamefont{{Guth}}(1981)}]{Guth-1981}
\bibinfo{author}{\bibfnamefont{A.~H.} \bibnamefont{{Guth}}},
  \bibinfo{journal}{\prd} \textbf{\bibinfo{volume}{23}}, \bibinfo{pages}{347}
  (\bibinfo{year}{1981}).

\bibitem[{\citenamefont{{Liddle}}(1999)}]{Liddle-1999mq}
\bibinfo{author}{\bibfnamefont{A.~R.} \bibnamefont{{Liddle}}}, in
  \emph{\bibinfo{booktitle}{Trieste 1998, High energy physics and cosmology}}
  (\bibinfo{organization}{Abdus Salam International Centre for Theoretical
  Physics}, \bibinfo{address}{Trieste, Italy}, \bibinfo{year}{1999}), pp.
  \bibinfo{pages}{260--295}, \eprint{astro-ph/9901124}.

\bibitem[{\citenamefont{Ghosh et~al.}(2000)\citenamefont{Ghosh, Madden, and
  Veneziano}}]{Ghosh-1999xm}
\bibinfo{author}{\bibfnamefont{A.}~\bibnamefont{Ghosh}},
  \bibinfo{author}{\bibfnamefont{R.}~\bibnamefont{Madden}}, \bibnamefont{and}
  \bibinfo{author}{\bibfnamefont{G.}~\bibnamefont{Veneziano}},
  \bibinfo{journal}{Nucl. Phys.} \textbf{\bibinfo{volume}{B570}},
  \bibinfo{pages}{207} (\bibinfo{year}{2000}), \eprint{hep-th/9908024}.

\end{thebibliography}

\newpage


%

\begin{table}
\caption{\label{tab:shotdata}Shot Selection}
\begin{ruledtabular}
\begin{tabular}{rrcl}
Mode & Number & Time(ms) & Description \\
L & 98777 & 1600 & L-mode \\
L-ITB & 102942 & 1400 & L-mode with Internal Transport Barrier \\
H & 99411 & 1800 & H-mode \\
QH & 122338 & 2750 & Quiescent H-mode
\end{tabular}
\end{ruledtabular}
\end{table}

\clearpage
\newpage

\begin{table}
\caption{\label{tab:tautab}Pseudoplasticity and Confinement Quality}
\begin{ruledtabular}
\begin{tabular}{rrccl}
Mode & Number & $S_\vphi$ & $S_\vtheta$ & Q \\
L & 98777 & -3.73(12) & -2.31(11) & 6.04 \\
L-ITB & 102942 & -3.25(13) & -1.93(06) & 5.18 \\
H & 99411 & -2.28(09) & -1.66(04) & 3.94 \\
QH & 122338 & -2.09(07) & -1.32(04) & 3.41
\end{tabular}
\end{ruledtabular}
\end{table}

\clearpage
\newpage

\begin{enumerate}
\item Figure 1. 98777 Velocity and viscosity profiles for the low viscosity solution.
\item Figure 2. 98777 Velocity and viscosity profiles for the high viscosity solution.
\item Figure 3. 98777 Poloidal coefficient profiles for the low viscosity solution.
\item Figure 4. 98777 Poloidal coefficient profiles for the high viscosity solution.
\item Figure 5. 98777 Vanishing coefficient profiles for both the old and new high viscosity solutions.
\item Figure 6. 99411 Velocity and viscosity profiles for the high viscosity solution.
\item Figure 7. 102942 Velocity and viscosity profiles for the high viscosity solution.
\item Figure 8. 122338 Velocity and viscosity profiles for the high viscosity solution.
\item Figure 9. Viscosity {\it versus} velocity for all confinement regimes.
\item Figure 10. Collisional viscosity {\it versus} gyroviscosity for all confinement regimes.
\item Figure 11. Radial electric field profiles for all confinement modes.
\item Figure 12. 98777 Input profiles for poloidal magnetic field, toroidal electric field, toroidal momentum injection, C6 toroidal velocity, C6 poloidal velocity, temperature, and density.
\item Figure 13. 99411 Input profiles for poloidal magnetic field, toroidal electric field, toroidal momentum injection, C6 toroidal velocity, C6 poloidal velocity, temperature, and density.
\item Figure 14. 102942 Input profiles for poloidal magnetic field, toroidal electric field, toroidal momentum injection, C6 toroidal velocity, C6 poloidal velocity, temperature, and density.
\item Figure 15. 122338 Input profiles for poloidal magnetic field, toroidal electric field, toroidal momentum injection, C6 toroidal velocity, C6 poloidal velocity, temperature, and density.
\item Figure 16. 99411 Poloidal coefficient profiles for the high viscosity solution.
\item Figure 17. 102942 Poloidal coefficient profiles for the high viscosity solution.
\item Figure 18. 122338 Poloidal coefficient profiles for the high viscosity solution.
\end{enumerate}

\clearpage
\newpage

\begin{figure}%
\includegraphics{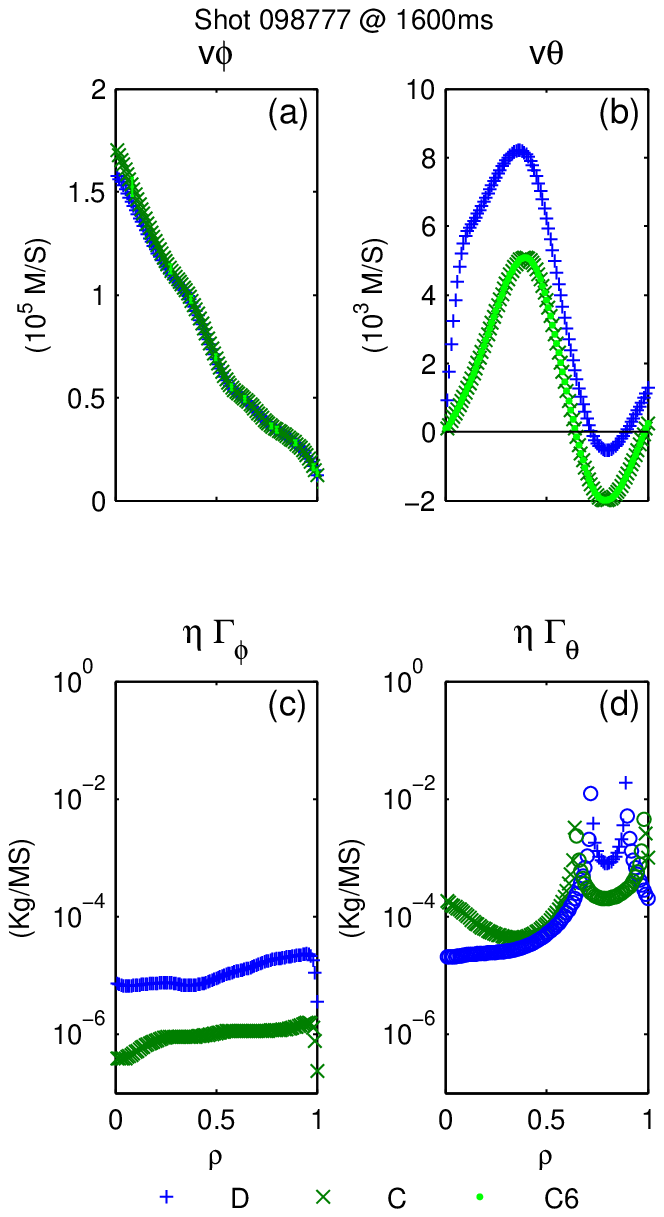}
\caption[98777 Velocity and viscosity profiles.]{\label{fig:Llov}(Color online).  98777 Velocity and viscosity profiles for the low viscosity solution.}
\end{figure}

\clearpage

\begin{figure}%
\includegraphics{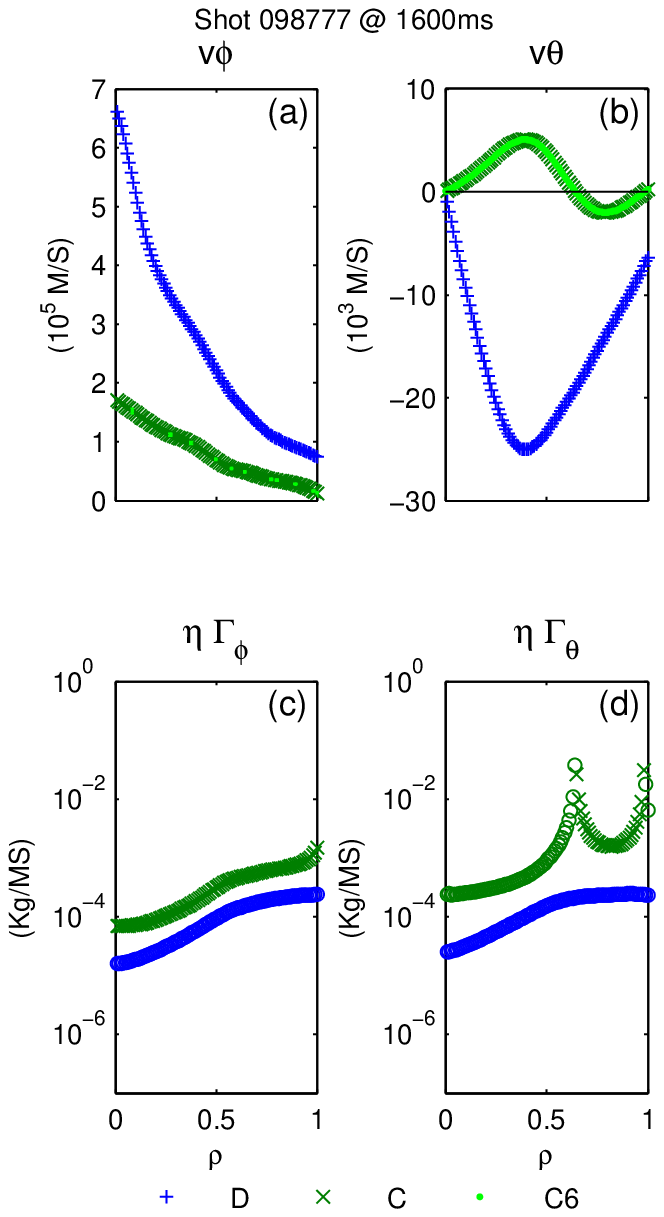}
\caption[98777 Velocity and viscosity profiles.]{\label{fig:Lhiv}(Color online).  98777 Velocity and viscosity profiles for the high viscosity solution.}
\end{figure}

\clearpage

\begin{figure}%
\includegraphics{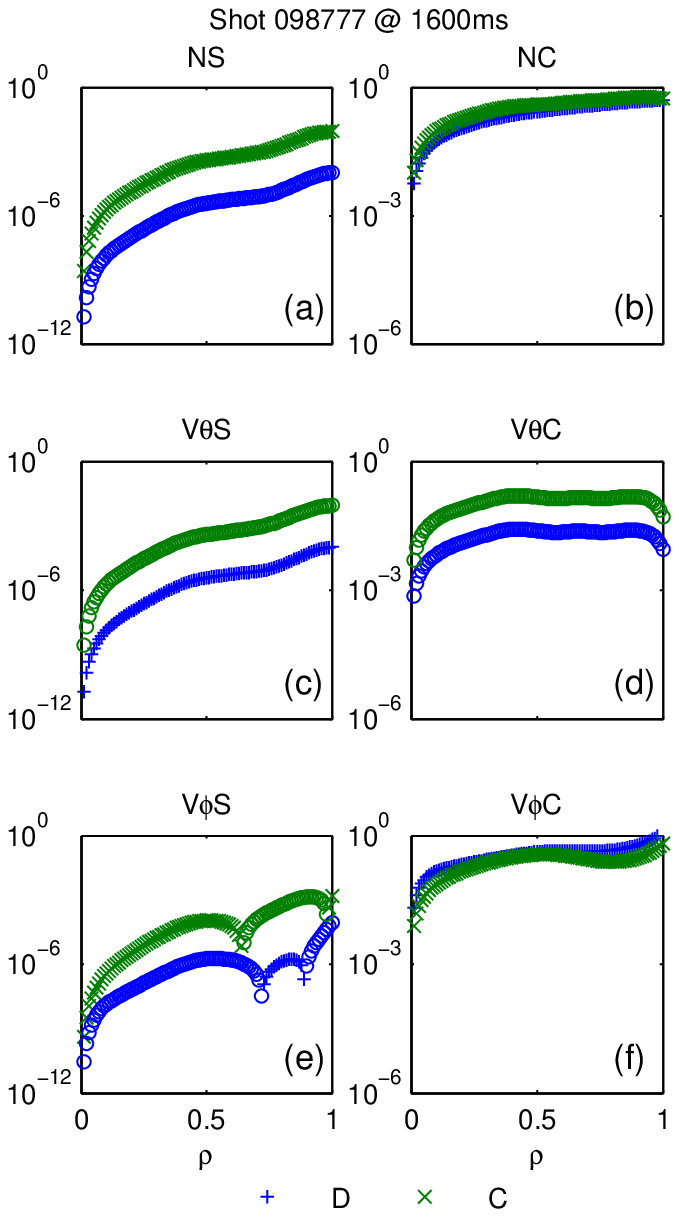}
\caption[98777 Poloidal coefficient profiles.]{\label{fig:Llop}(Color online).  98777 Poloidal coefficient profiles for the low viscosity solution.}
\end{figure}

\clearpage

\begin{figure}%
\includegraphics{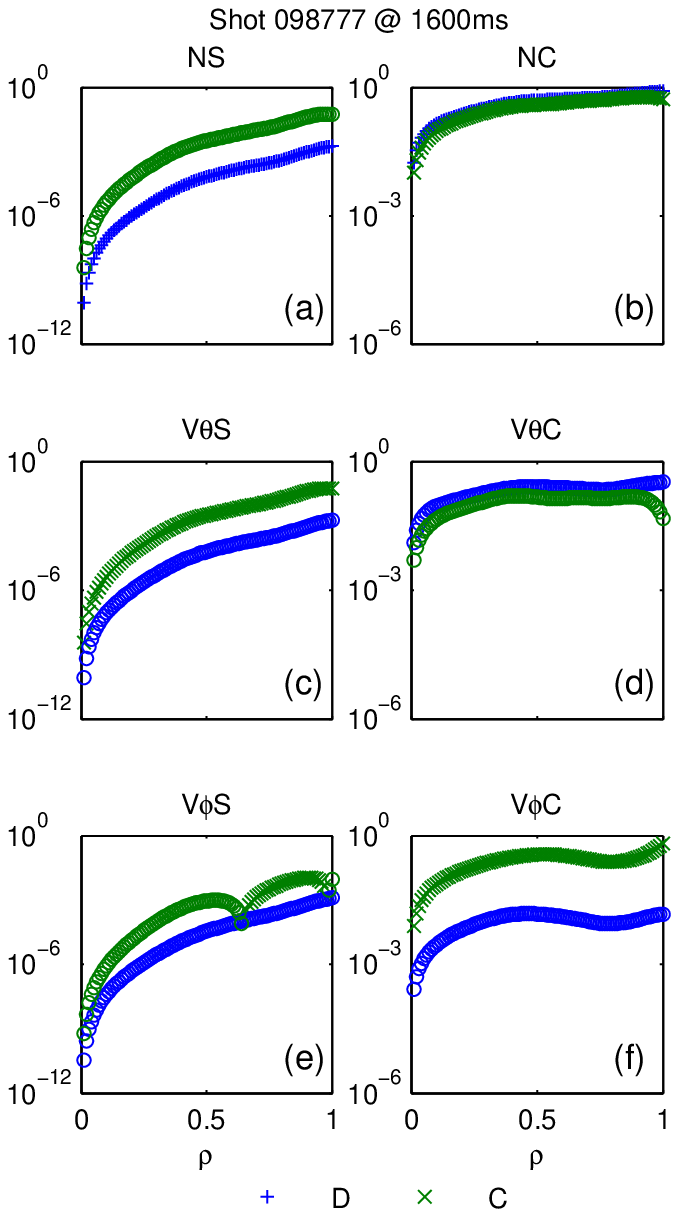}
\caption[98777 Poloidal coefficient profiles.]{\label{fig:Lhip}(Color online).  98777 Poloidal coefficient profiles for the high viscosity solution.}
\end{figure}

\clearpage

\begin{figure}%
\includegraphics{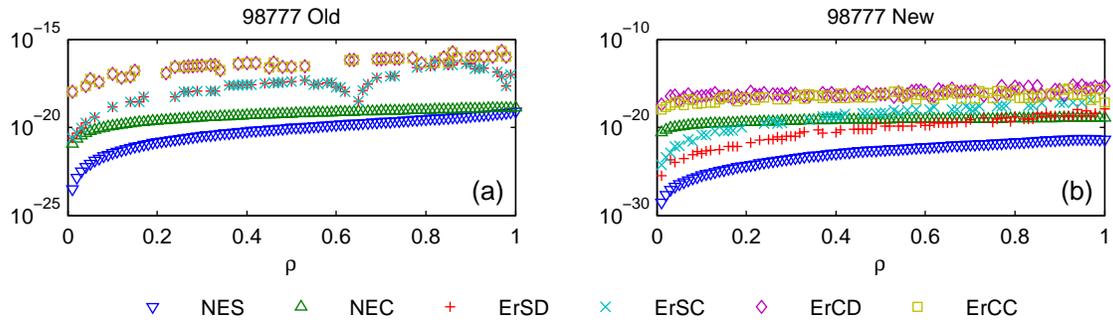}
\caption[99777 Vanishing coefficient profiles.]{\label{fig:ernecs}(Color online).  98777 Vanishing coefficient profiles for both the old and new high viscosity solutions.}
\end{figure}

\clearpage

\begin{figure}%
\includegraphics{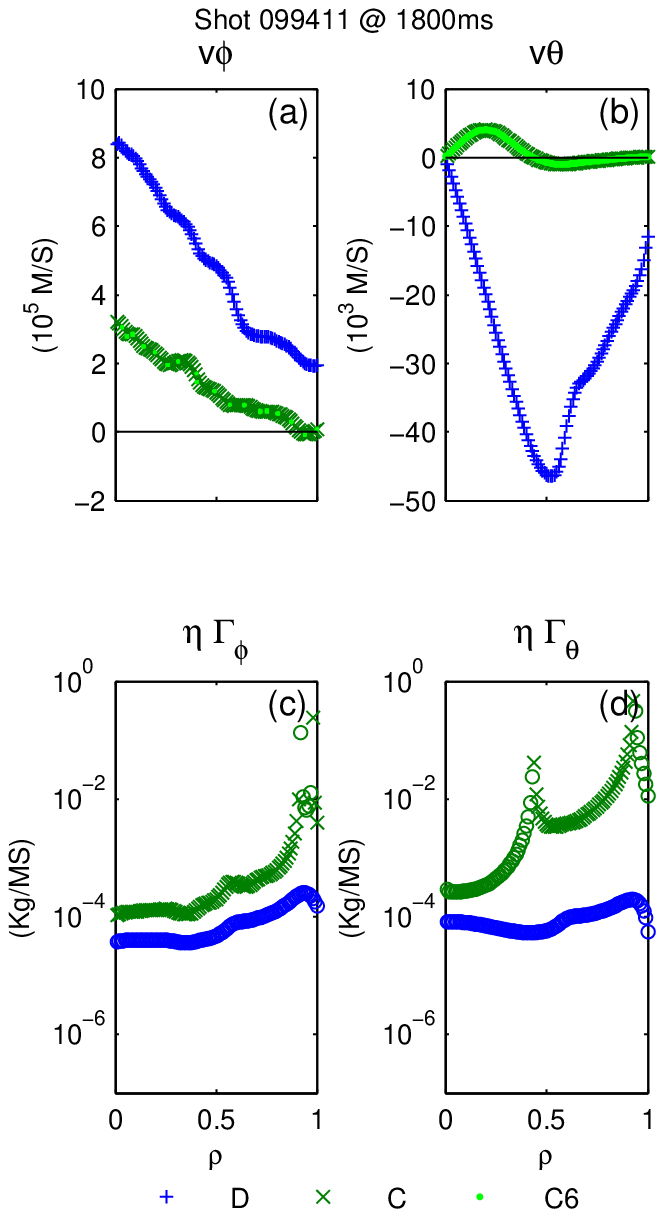}
\caption[99411 Velocity and viscosity profiles.]{\label{fig:Hv}(Color online).  99411 Velocity and viscosity profiles for the high viscosity solution.}
\end{figure}

\clearpage

\begin{figure}%
\includegraphics{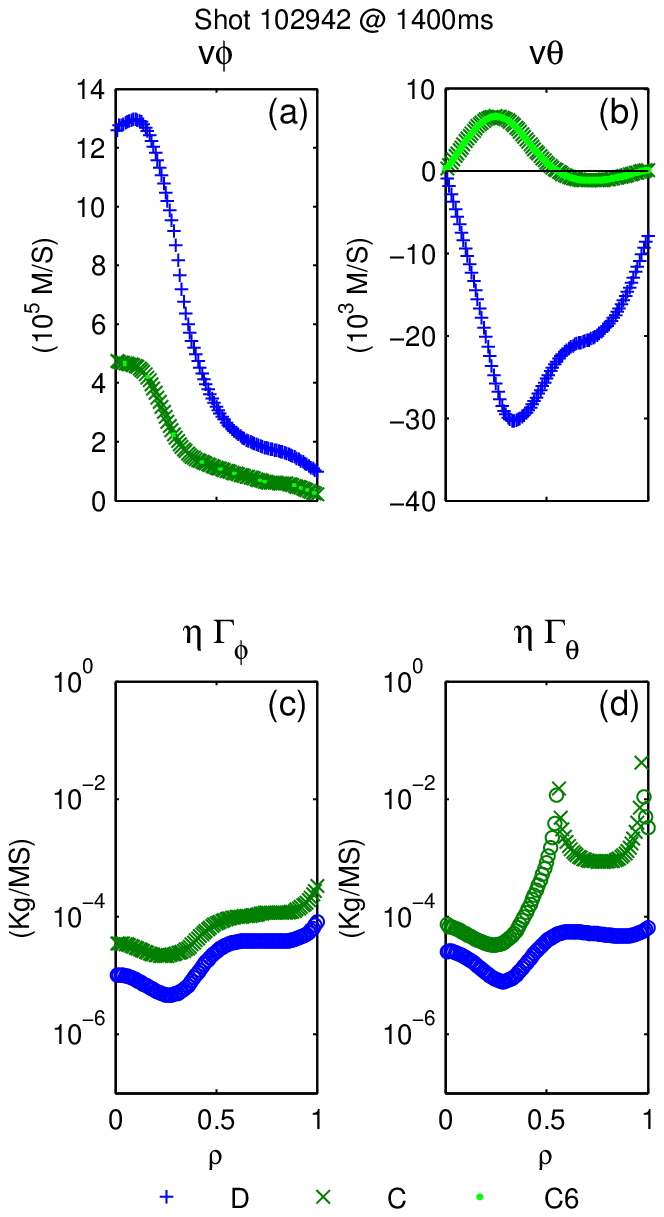}
\caption[102942 Velocity and viscosity profiles.]{\label{fig:LITBv}(Color online).  102942 Velocity and viscosity profiles for the high viscosity solution.}
\end{figure}

\clearpage

\begin{figure}%
\includegraphics{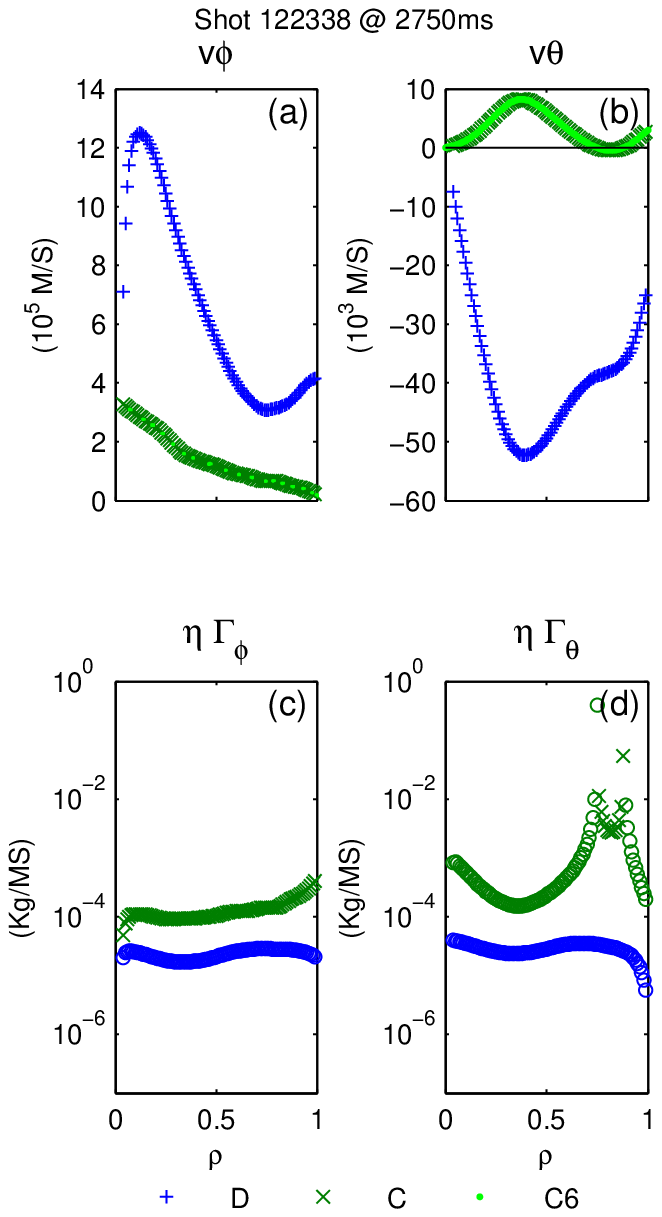}
\caption[122338 Velocity and viscosity profiles.]{\label{fig:QHv}(Color online).  122338 Velocity and viscosity profiles for the high viscosity solution.}
\end{figure}

\clearpage

\begin{figure}%
\includegraphics{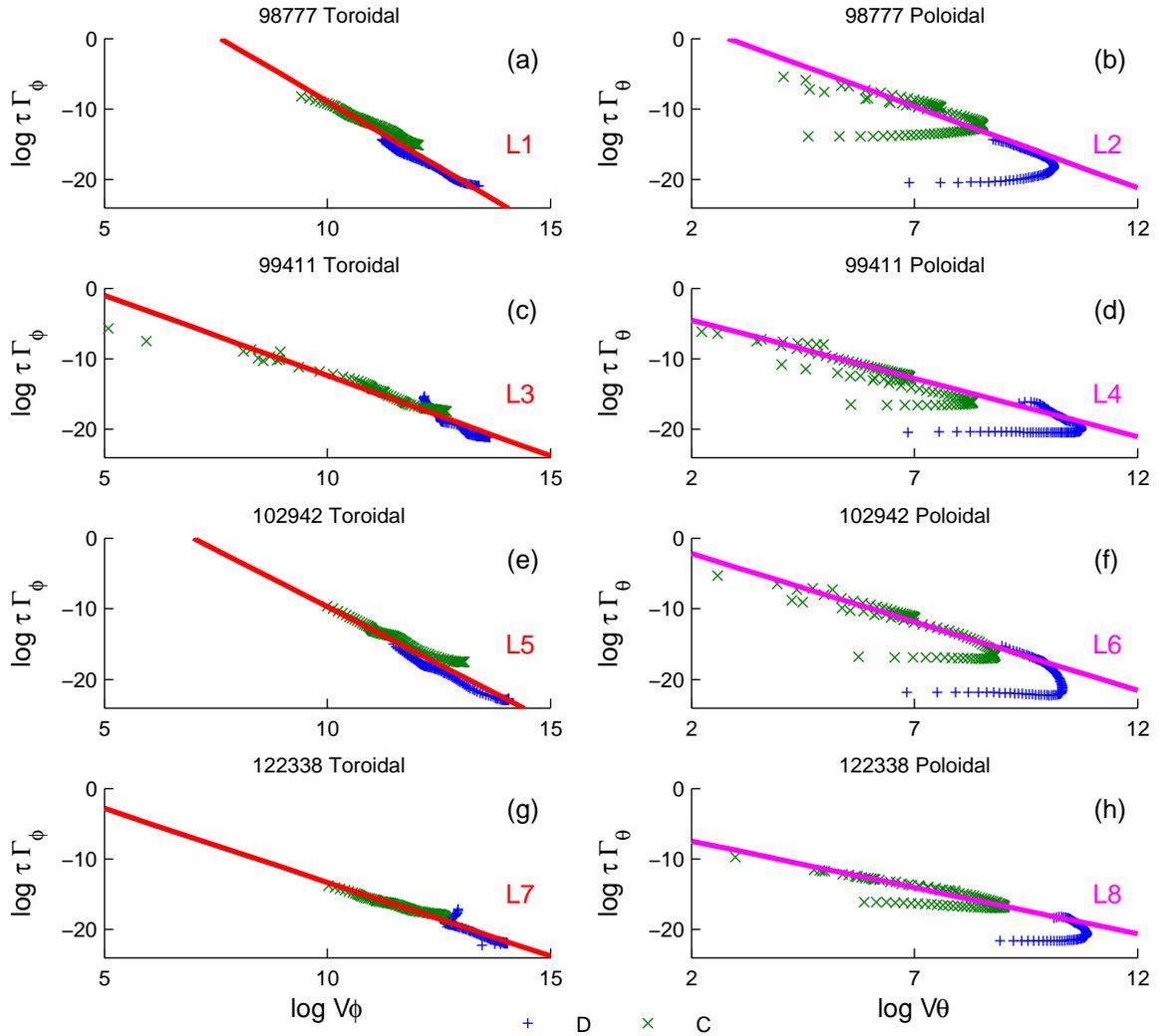}
\caption[Viscosity {\it versus} velocity.]{\label{fig:tauvs}(Color online).  Viscosity {\it versus} velocity for all confinement regimes.}
\end{figure}

\clearpage

\begin{figure}%
\includegraphics{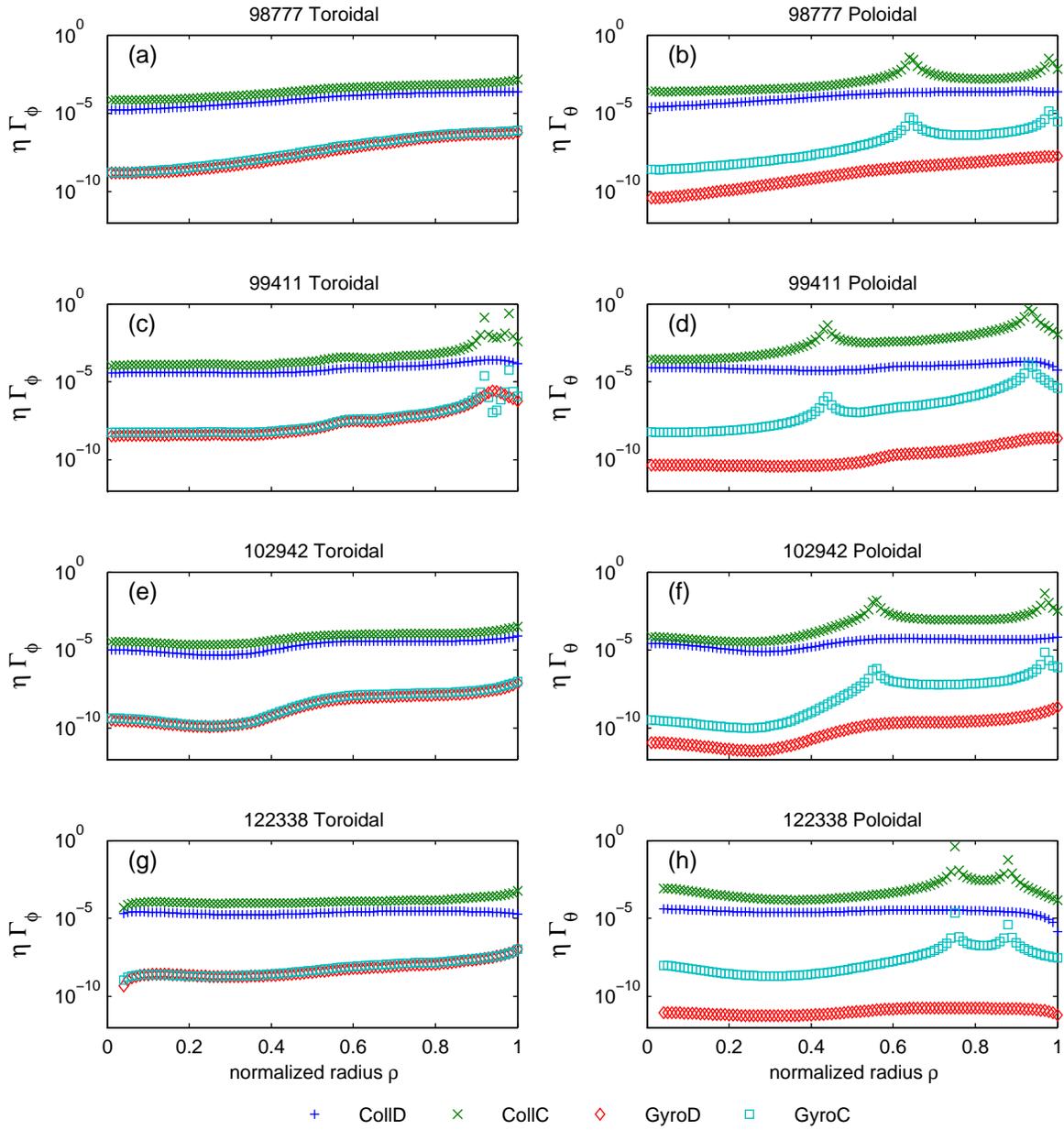}
\caption[Collisional viscosity {\it versus} gyroviscosity.]{\label{fig:viscs}(Color online).  Collisional viscosity {\it versus} gyroviscosity for all confinement regimes.}
\end{figure}

\clearpage

\begin{figure}%
\includegraphics{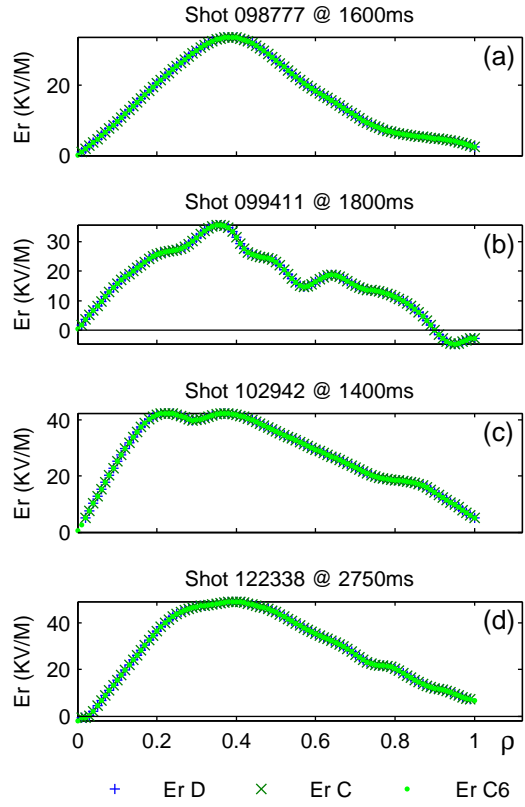}
\caption[Radial electric field profiles.]{\label{fig:Lerad}(Color online).  Radial electric field profiles for all confinement modes.}
\end{figure}

\clearpage

\begin{figure}%
\includegraphics{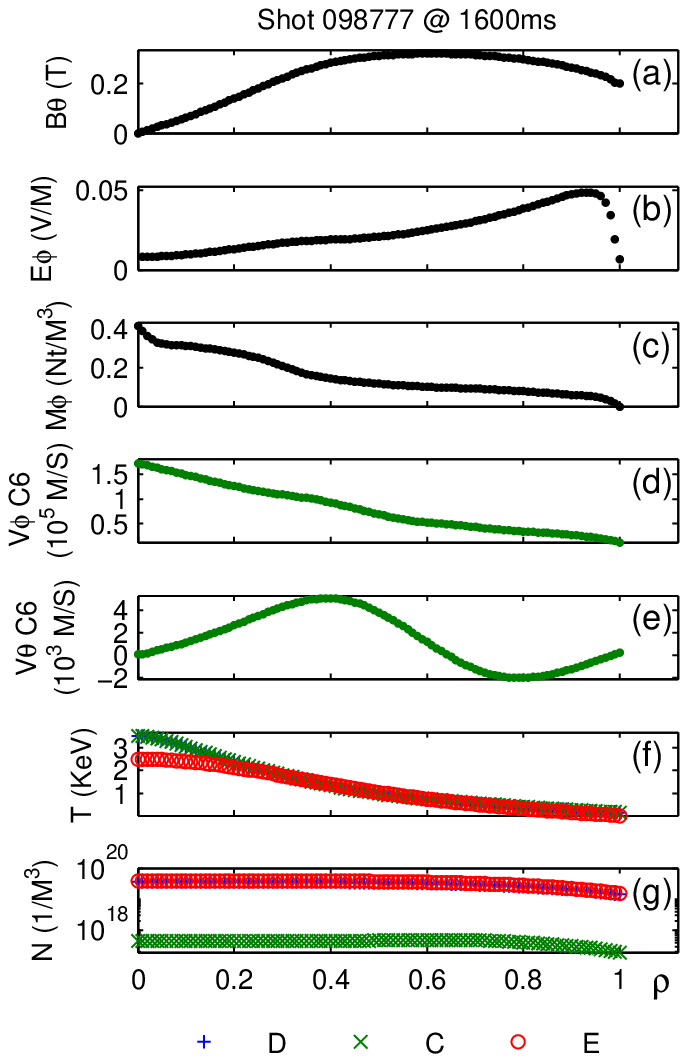}
\caption[98777 Input profiles.]{\label{fig:Lin}(Color online).  98777 Input profiles for poloidal magnetic field, toroidal electric field, toroidal momentum injection, C6 toroidal velocity, C6 poloidal velocity, temperature, and density.}
\end{figure}

\clearpage

\begin{figure}%
\includegraphics{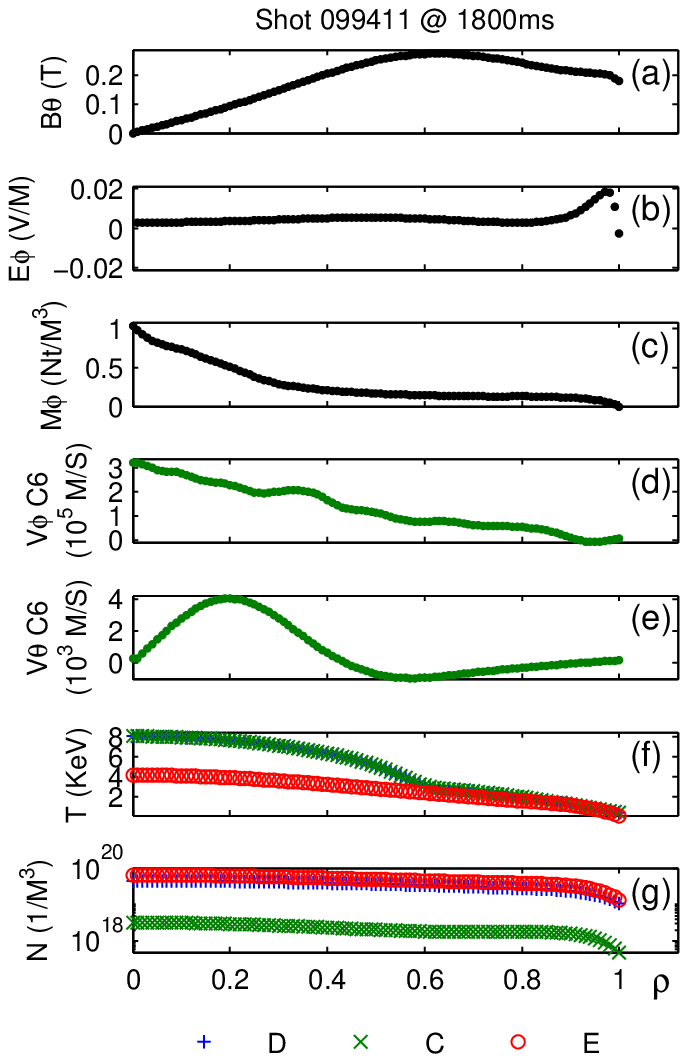}
\caption[99411 Input profiles.]{\label{fig:Hin}(Color online).  99411 Input profiles for poloidal magnetic field, toroidal electric field, toroidal momentum injection, C6 toroidal velocity, C6 poloidal velocity, temperature, and density.}
\end{figure}

\clearpage

\begin{figure}%
\includegraphics{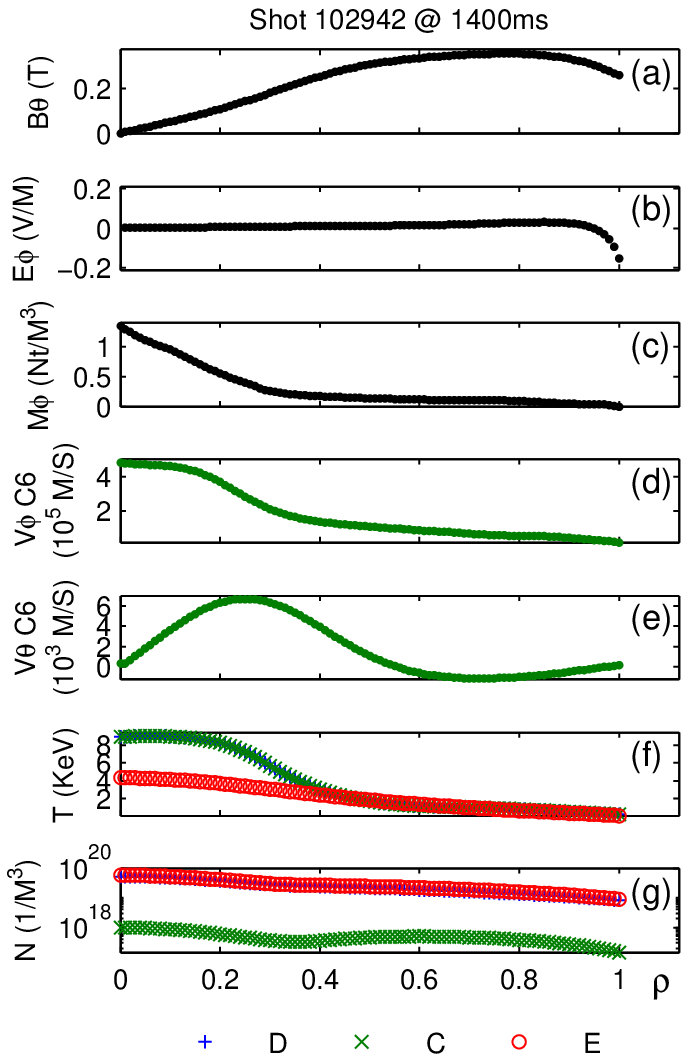}
\caption[102942 Input profiles.]{\label{fig:LITBin}(Color online).  102942 Input profiles for poloidal magnetic field, toroidal electric field, toroidal momentum injection, C6 toroidal velocity, C6 poloidal velocity, temperature, and density.}
\end{figure}

\clearpage

\begin{figure}%
\includegraphics{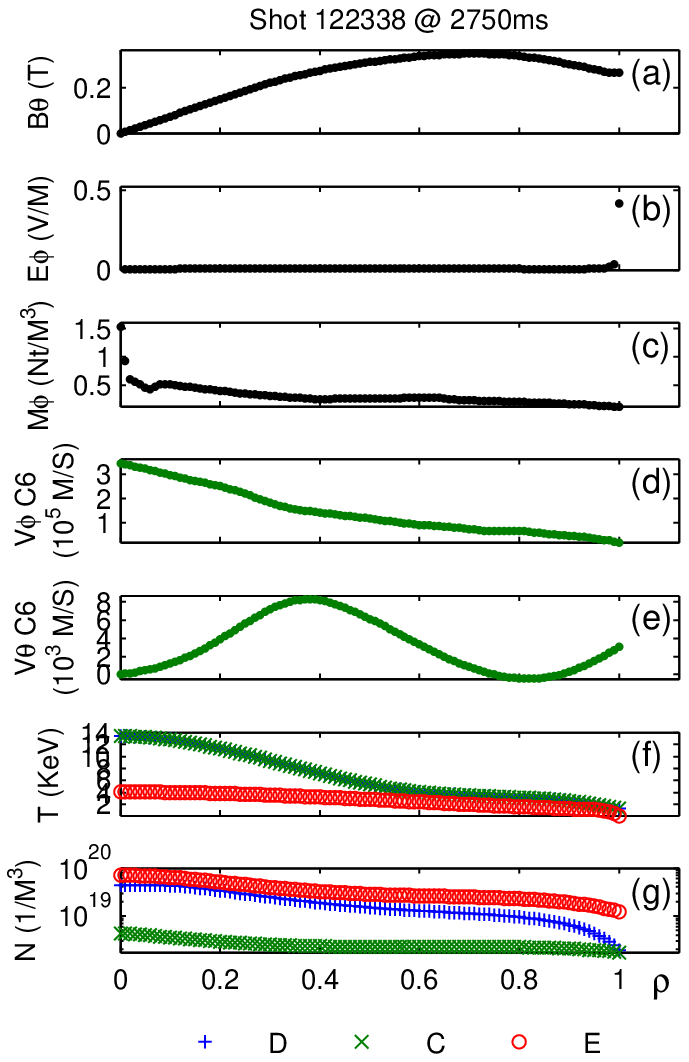}
\caption[122338 Input profiles.]{\label{fig:QHin}(Color online).  122338 Input profiles for poloidal magnetic field, toroidal electric field, toroidal momentum injection, C6 toroidal velocity, C6 poloidal velocity, temperature, and density.}
\end{figure}

\clearpage

\begin{figure}%
\includegraphics{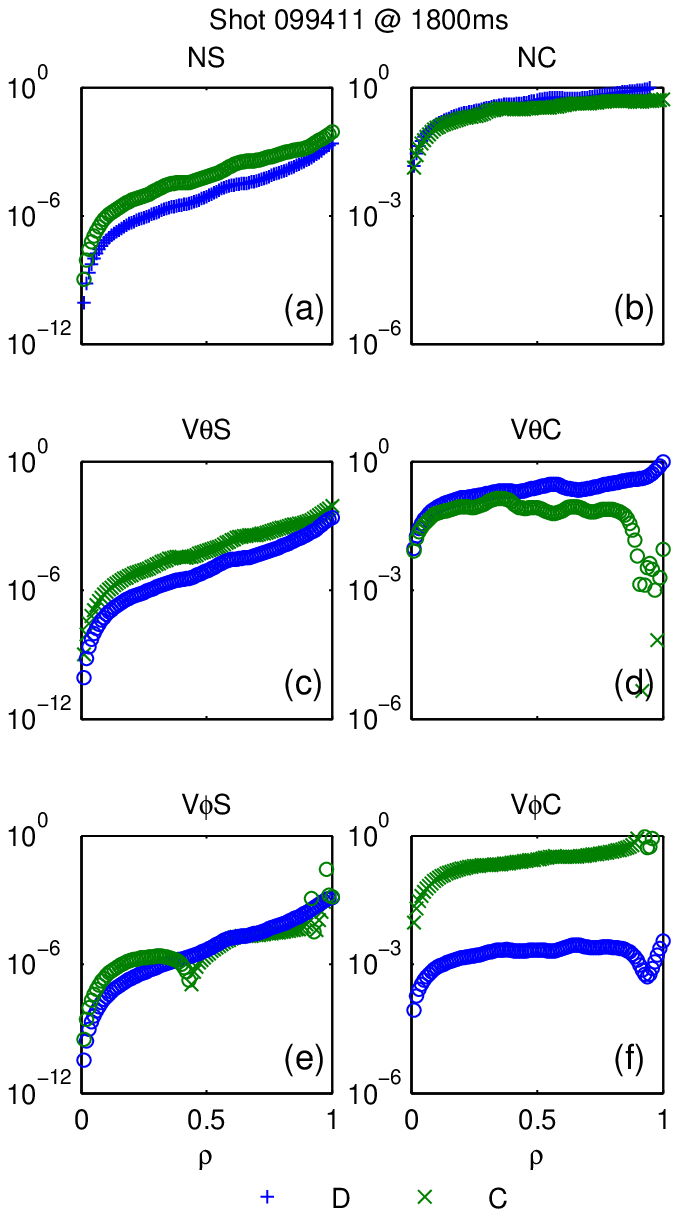}
\caption[99411 Poloidal coefficient profiles.]{\label{fig:Hp}(Color online).  99411 Poloidal coefficient profiles for the high viscosity solution.}
\end{figure}

\clearpage

\begin{figure}%
\includegraphics{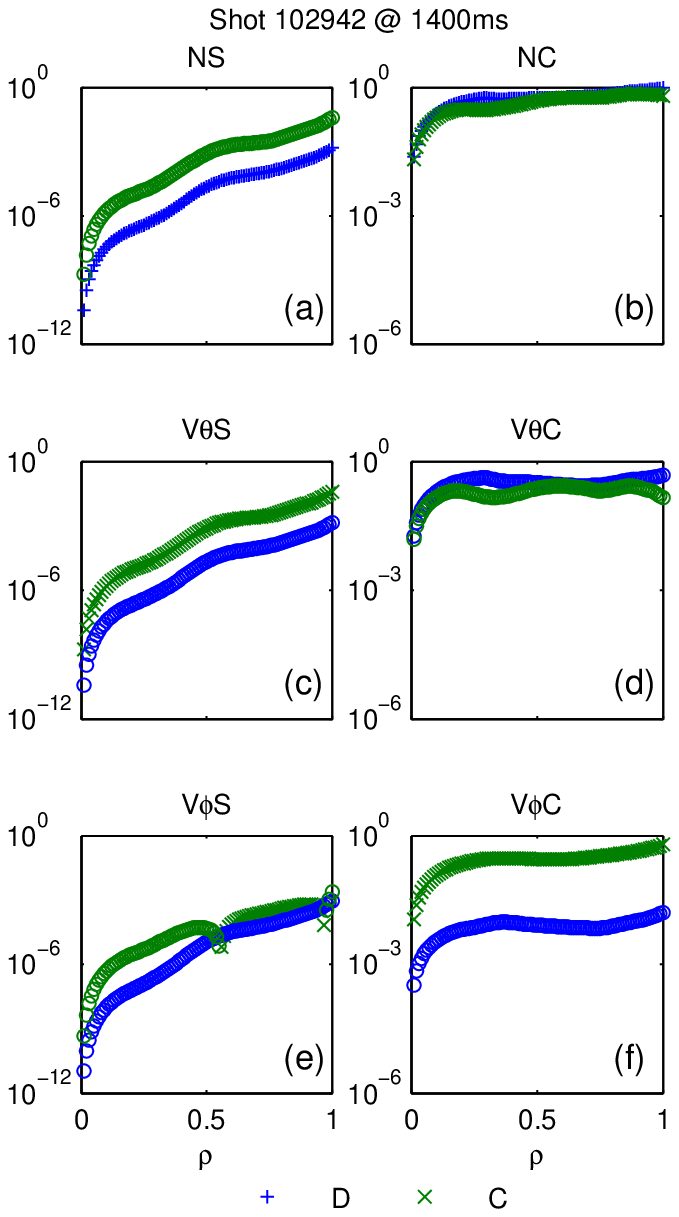}
\caption[102942 Poloidal coefficient profiles.]{\label{fig:LITBp}(Color online).  102942 Poloidal coefficient profiles for the high viscosity solution.}
\end{figure}

\clearpage

\begin{figure}%
\includegraphics{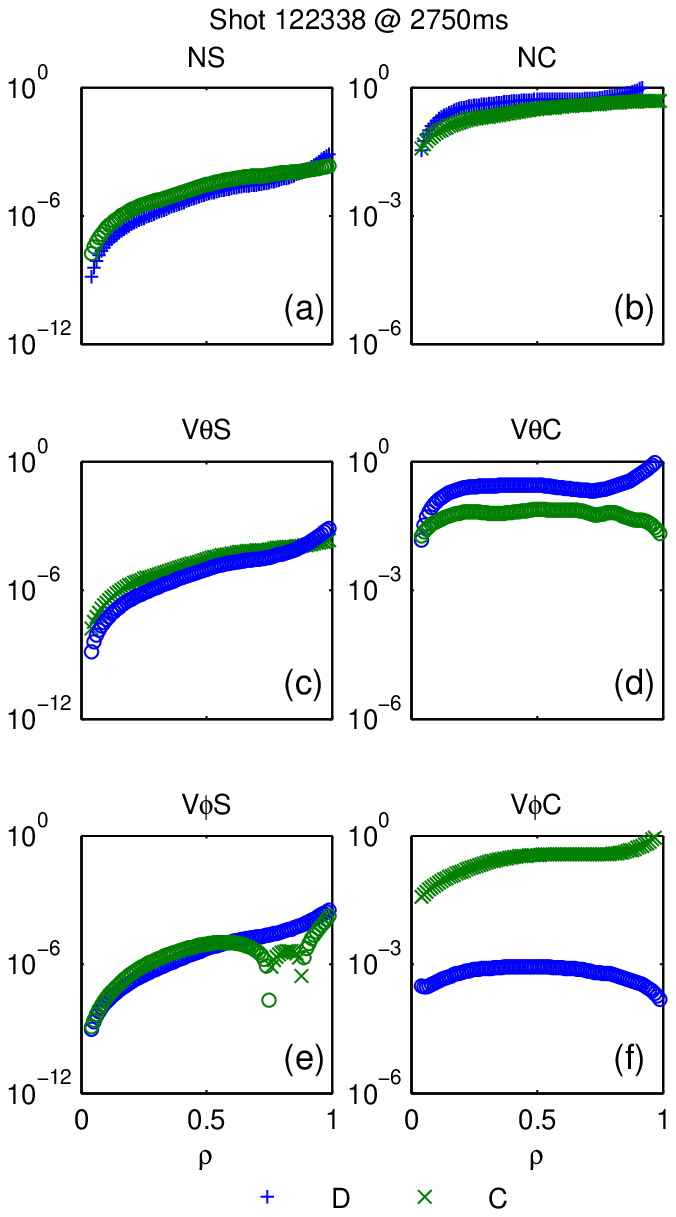}
\caption[122338 Poloidal coefficient profiles.]{\label{fig:QHp}(Color online).  122338 Poloidal coefficient profiles for the high viscosity solution.}
\end{figure}

\clearpage

\end{document}